\documentclass[12pt]{article}
\usepackage{epsf,amsmath,euscript}

\renewcommand{\arraystretch}{1.2}

\catcode`@=11
\def\@citex[#1]#2{%
  \if@filesw\immediate\write\@auxout{\string\citation{#2}}\fi
  \def\@citea{}\@cite{\@for\@citeb:=#2\do
    {\@citea\def\@citea{,\penalty\@m}\@ifundefined
      {b@\@citeb}{{\bf ?}\@warning
{Citation `\@citeb' on page \thepage \space undefined}}%
      \hbox{\csname b@\@citeb\endcsname}}}{#1}}
\def\citer{\@ifnextchar [{\@tempswatrue\@citexr}{\@tempswafalse\@citexr[]}}
  \def\@citexr[#1]#2{%
    \if@filesw\immediate\write\@auxout{\string\citation{#2}}\fi
    \def\@citea{}\@cite{\@for\@citeb:=#2\do
      {\@citea\def\@citea{--\penalty\@m}\@ifundefined
{b@\@citeb}{{\bf ?}\@warning
{Citation `\@citeb' on page \thepage \space undefined}}%
\hbox{\csname b@\@citeb\endcsname}}}{#1}}

\newif\ifpdf\ifx\pdfoutput\undefined\pdffalse\else\pdfoutput=1\pdftrue\fi
\newcommand{\pdfgraphics}{\ifpdf\DeclareGraphicsExtensions{.pdf, .jpg}\else\DeclareGraphicsExtensions{.eps, .ps, .jpg}\fi}
\usepackage{graphicx}

\begin{document}

\pdfgraphics

\begin{titlepage}

\begin{flushright}
CLNS~04/1870\\
LMU~03/04\\
%{\tt hep-ph/0408231}\\[0.2cm]
August 2004
\end{flushright}

\vspace{0.7cm}
\begin{center}
\Large\bf 
Constraining the Unitarity Triangle\\
with $B\to V\gamma$
\end{center}

\vspace{0.8cm}
\begin{center}
{\sc Stefan~W.~Bosch$^a$ and Gerhard Buchalla$^b$}\\
\vspace{0.7cm}
{\sl ${}^a$Institute for High-Energy Phenomenology\\ 
Newman Laboratory for Elementary-Particle Physics, 
Cornell University\\
Ithaca, NY 14853, U.S.A.}\\
\vspace{0.3cm}
{\sl ${}^b$ Ludwig-Maximilians-Universit\"at M\"unchen, 
Department f\"ur Physik\\
Theresienstra\ss e 37, D-80333 Munich, Germany}
\end{center}

\vspace{1.0cm}
\begin{abstract}
\vspace{0.2cm}\noindent
We discuss the exclusive radiative decays $B\to K^{*}\gamma$, 
$B \to\rho\gamma$, and $B\to\omega\gamma$ in QCD factorization within the 
Standard Model. The analysis is based on the heavy-quark limit of QCD. 
Our results for these decays are complete to next-to-leading order in QCD and 
to leading order in the heavy-quark limit. Special emphasis 
is placed on constraining the CKM-unitarity triangle from these observables. 
We propose a theoretically clean method to determine CKM parameters
from the ratio of the $B\to\rho l\nu$ decay spectrum to the branching
fraction of $B\to\rho\gamma$. The method is based on the cancellation
of soft hadronic form factors in the large energy limit, which occurs
in a suitable region of phase space.
The ratio of the $B\to\rho\gamma$ and $B\to K^{*}\gamma$ branching fractions 
determines the side $R_{t}$ of the standard unitarity triangle with
reduced hadronic uncertainties. 
The recent Babar bound on $B(B^0\to\rho^0\gamma)$ implies
$R_t < 0.81\, (\xi/1.3)$, with the limiting uncertainty coming only 
from the SU(3) breaking form factor ratio $\xi$.
This constraint is already getting competitive with the constraint from 
$B_{s}$-$\bar B_{s}$ mixing. Phenomenological implications from 
isospin-breaking effects are briefly discussed. 

\end{abstract}
\vfil

\end{titlepage}

\section{Introduction}
\label{intro}
The radiative transition $b\to s\gamma$ is one of the most 
important processes for the study of flavour physics.
As a flavour-changing neutral current interaction
it is a genuine quantum effect within the Standard Model (SM)
and has a high sensitivity to new dynamics at short-distance scales.
The cleanest way to probe $b\to s\gamma$ is the measurement of the
inclusive decay $B\to X_s\gamma$, where the impact of strong
interactions is well under control (see \cite{review} for a recent review).
 For exclusive channels such as
$B\to K^*\gamma$, which depend on hadronic quantities describing
the hadronization of the final state quarks into
a single $K^*$, a theoretical treatment is more difficult.
At present, the decay $B\to X_s\gamma$ already yields strong tests
of the SM and valuable constraints on its possible extensions.

In contrast, not much is currently known experimentally about
$b\to d\gamma$ transitions, the Cabibbo-suppressed counterparts
of $b\to s\gamma$. They depend on the less well determined
weak mixing parameter $V_{td}$, rather than $V_{ts}$,
and could be differently affected by new physics. For these reasons
a measurement of $b\to d\gamma$ will be very important.
However, the inclusive measurement of $B\to X_d\gamma$, theoretically
prefered, appears almost impossible because of the dominating background
from $B\to X_s\gamma$. Therefore, exclusive channels such as
$B\to\rho\gamma$ and $B\to\omega\gamma$ become the only way to access 
$b\to d\gamma$ transitions in the foreseeable future.

The CP-averaged branching ratios of exclusive radiative channels are measured 
to be \cite{BKgamexp} 
\begin{eqnarray}
B(B^{0} \to K^{*0}\gamma) &=& (4.01 \pm 0.20) \cdot 10^{-5} \label{b0kg}\\
B(B^{+} \to K^{*+}\gamma) &=& (4.03 \pm 0.26) \cdot 10^{-5} \label{bpkg}
\end{eqnarray}
and bounded with 90\% confidence level by Babar as \cite{Brhoexp1} 
\begin{eqnarray}
B(B^{0} \to \omega^{0}\gamma) &<& 1.0 \cdot 10^{-6} \label{b0og}\\
B(B^{0} \to \rho^{0}\gamma)        &<& 0.4 \cdot 10^{-6} \label{b0rg}\\
B(B^{+} \to \rho^{+}\gamma)        &<& 1.8 \cdot 10^{-6} \label{bprg}
\end{eqnarray}
The corresponding results from Belle read \cite{Brhoexp2}
\begin{eqnarray}
B(B^{0} \to \omega^{0}\gamma) &<& 0.8 \cdot 10^{-6} \label{b0og2}\\
B(B^{0} \to \rho^{0}\gamma)        &<& 0.8 \cdot 10^{-6} \label{b0rg2}\\
B(B^{+} \to \rho^{+}\gamma)        &<& 2.2 \cdot 10^{-6} \label{bprg2}
\end{eqnarray}

Even though a theoretical treatment of the exclusive decays
$B\to\rho\gamma$, $B\to\omega\gamma$, and $B\to K^*\gamma$ is more challenging 
than of the inclusive modes, there are circumstances that help us
to make this task tractable and that will eventually yield
useful phenomenological results.
First, recent studies of exclusive hadronic modes in the heavy-quark limit
have led to a better understanding of the strong dynamics
of these decays \cite{BBNS,BBVgam,BFS} by establishing factorization
formulas in QCD. For the decays $B\to V\gamma$ \cite{BBVgam,BFS} this
approach resulted in particular in a calculation of light-quark
loop amplitudes that before constituted an uncontrollable source
of uncertainty. In addition it became possible to extend the computation
of $B\to V\gamma$ amplitudes systematically to next-to-leading
order (NLO) in QCD \cite{BBVgam,BFS,AP}, improving on previous analyses 
\cite{AAWGSW,DLTES}.
Second, the impact of hadronic form factors, which dominates
theoretical uncertainties, can be reduced by taking the ratio
$B(B\to\rho\gamma)/B(B\to K^*\gamma)$. The ratio of the corresponding
form factors is equal to unity in the limit of $SU(3)$-flavour symmetry
and the hadronic uncertainty is reduced to the effect of
$SU(3)$ breaking, which still needs to be estimated. 
Furthermore, the ratio of the $B\to\rho\gamma$ and $B\to K^{*}\gamma$ 
branching fractions is, at leading order in $\alpha_{s}$, directly 
proportional to the side $R_{t}$ in the standard unitarity triangle (UT), where
\begin{equation}\label{Rtdef}
R_t\equiv\sqrt{(1-\bar\rho)^2+\bar\eta^2}=
\frac{1}{\lambda}\left|\frac{V_{td}}{V_{ts}}\right|
\end{equation}
Here $\lambda$, $\bar\rho$, and $\bar\eta$ are Wolfenstein
parameters. 
Having the complete NLO result for the decay amplitudes in $B\to V\gamma$ 
at hand, we can calculate $\alpha_{s}$ corrections to their relation 
with $R_t$ and evaluate the implications in the ($\bar\rho,\bar\eta)$ plane 
\cite{AliLunghi,HL,UTBVgam}.

Another possibility to reduce hadronic uncertainties consists in
taking the ratio of $B\to\rho l\nu$ and $B\to\rho\gamma$ branching
fractions. Using relations between the $B\to\rho$ form factors in
the large energy linit, it can be shown that this ratio is free of
long-distance QCD effects in a certain region of $B\to\rho l\nu$
phase space. The form factors cancel in this situation, up to
calculable ${\cal O}(\alpha_s)$ corrections, which leads to a 
model-independent relationship of $B\to\rho l\nu$ and $B\to\rho\gamma$
observables to the CKM quantity
\begin{equation}
\left|\frac{V_{ud}V_{ub}}{V_{td}V_{tb}}\right|^2 =
\frac{\bar\rho^2 +\bar\eta^2}{(1-\bar\rho)^2+\bar\eta^2}
\end{equation}

It is the purpose of this paper to investigate how $B\to V\gamma$ decays
can be used to constrain the parameters of the unitarity triangle.
Such constraints simultaneously provide a test for new physics.
The various sources of
uncertainty will be discussed in detail in order to quantify the
potential of these important decays.
In section 2 we recall the analysis of $B\to V\gamma$ decays at
next-to-leading order within the framework of factorization
in the heavy-quark limit. The extraction of CKM parameters based on the
ratios $B(B\to\rho\gamma)/B(B\to K^*\gamma)$ is the subject of
section 3.
In section 4 we discuss how theoretically clean information on
CKM quantities can be obtained from combining a measurement of
$B(B\to\rho^0\gamma)$ with a Dalitz-plot analysis of $B\to\rho l\nu$
decays. Section 5 contains an update on observables of isospin breaking
in $B\to V\gamma$ and section 6 is devoted to a discussion of the
decay mode $B\to\omega\gamma$. We present our conclusions
in section 7.

\section{$B\to V\gamma$ at NLO in QCD}
\label{sec:BVgamNLO}
Let us briefly summarize the basic formulas relevant for the analysis of 
$B\to V\gamma$ at next-to-leading order in QCD. For more details we refer the 
reader to \cite{BBVgam,thesis}. The effective weak Hamiltonian for 
$b\to s\gamma$ transitions is
\begin{equation}\label{heff}
  {\cal H}_{eff}=\frac{G_F}{\sqrt{2}}\sum_{p=u,c}\lambda_p^{(s)}
\bigg( \sum_{i=1}^2 C_i Q^p_i +\sum_{j=3}^8 C_j Q_j\bigg)
\end{equation}
where
\begin{equation}\label{lamps}
\lambda_p^{(s)}=V^*_{ps}V_{pb}
\end{equation}
The relevant operators are the current-current operators $Q^p_{1,2}$, 
the QCD-penguin operators $Q_{3\ldots 6}$, and the electro- and chromomagnetic 
penguin operators $Q_{7,8}$.
The most important contributions come from $Q^p_{1,2}$ and $Q_{7,8}$,
which read
\begin{equation}\label{q1def}
Q^p_1 = (\bar sp)_{V-A}(\bar pb)_{V-A} \qquad
Q^p_2 = (\bar s_i p_j)_{V-A}(\bar p_j b_i)_{V-A} 
\end{equation}
\begin{equation}\label{q7def}
Q_7 = \frac{e}{8\pi^2}m_b\, 
        \bar s_i\sigma^{\mu\nu}(1+\gamma_5)b_i\, F_{\mu\nu}
\qquad
Q_8 = \frac{g}{8\pi^2}m_b\, 
        \bar s_i\sigma^{\mu\nu}(1+\gamma_5)T^a_{ij} b_j\, G^a_{\mu\nu}
\end{equation}
The impact of penguin operators $Q_3,\ldots Q_6$ is very small
for most applications, but will be included in the numerical
results presented below.
The effective Hamiltonian for $b\to d\gamma$ is obtained from 
(\ref{heff})--(\ref{q7def}) by the replacement $s\to d$. 

To evaluate the hadronic matrix elements of these operators we employ the heavy-quark limit 
$m_{b}\gg\Lambda_{\mathrm{QCD}}$ to get the factorization formula \cite{BBVgam,BFS}
\begin{equation}\label{fform}
  \langle V\gamma(\epsilon)|Q_i|\bar B\rangle =\Big[ F_{V} T^I_{i} + \!\int^1_0 \!\!d\xi\, dv \, T^{II}_i(\xi,v) \Phi_B(\xi) \Phi_V(v)\Big] \!\cdot\epsilon
\end{equation}
where $\epsilon$ is the photon polarization 4-vector. Here $F_{V}$ is a $B\to V$ transition form factor, and $\Phi_B$, $\Phi_V$ are leading-twist light-cone distribution amplitudes of the $B$ meson and the vector meson $V$, respectively. These quantities are universal, nonperturbative objects. They describe the long-distance dynamics of the matrix elements, which is factorized from the perturbative, short-distance interactions expressed in the hard-scattering kernels $T^I_{i}$ and $T^{II}_i$. The QCD factorization formula (\ref{fform}) holds up to corrections of relative order $\Lambda_{QCD}/m_b$.

To leading order in QCD and leading power in the heavy-quark limit, $Q_7$ gives the only contribution to the $B\to V\gamma$ amplitude. Its matrix element is simply expressed in terms of the standard form factor, $T^I_{7}$ is a purely kinematical function, and the spectator term $T^{II}_7$ is absent. At ${\cal O}(\alpha_s)$ the operators $Q_{1\ldots 6}$ and $Q_8$ start contributing and the factorization formula becomes nontrivial.

\begin{figure}
\center{\resizebox{0.75\textwidth}{!}{%
  \includegraphics{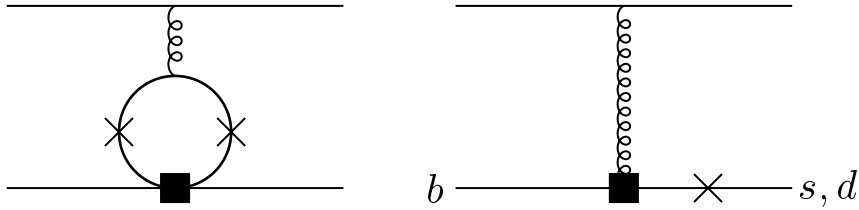}
}}
\caption[1]{\label{fig:qit2} \it ${\cal O}(\alpha_s)$ contribution at 
leading power to the hard-scattering kernels $T^{II}_i$ from four-quark 
operators $Q_i$ (left) and from $Q_8$. The crosses indicate the places 
where the emitted photon can be attached.}
\end{figure}

The relevant diagrams for the NLO hard-vertex corrections $T_i^I$ have been
computed in \cite{GHW,BCMU} to get the virtual corrections to the matrix 
elements for the inclusive $b\to s\gamma$ mode at next-to-leading order. 
For the exclusive modes the same corrections enter
the perturbative type I hard-scattering kernels. The non-vanishing 
contributions to $T^{II}_i$, where the spectator participates in the 
hard scattering, are shown in Fig.~\ref{fig:qit2}. 
%%%%%%%%%%%%%%%%%%%%%%%%%%%%%%%%%%%%%%%%%%%%%%%%%%%%%%%%%%%%%%%%%%%
%%%%%%%%%%%%%%%%%%%%%%%%%%%%%%%%%%%%%%%%%%%%%%%%%%%%%%%%%%%%%%%%%%%
We can express both the type I and type II contributions to the matrix 
elements $\langle Q_i\rangle$ in terms of the matrix element 
$\langle Q_7\rangle$, an explicit factor of $\alpha_s$, and hard-scattering 
functions $G_i$ and $H_i$, which are given in \cite{BBVgam,thesis}.

Weak annihilation contributions are suppressed by one power of 
$\Lambda_{\mathrm{QCD}}/m_{b}$ but nevertheless calculable in QCD 
factorization, because in the heavy-quark limit the colour-transparency 
argument applies to the emitted, highly energetic vector meson. 
Despite their suppression in $\Lambda_{\mathrm{QCD}}/m_{b}$, they can
be enhanced by large Wilson coefficients $C_{1,2}$ and thus still
give important corrections. This situation is relevant for $B\to\rho\gamma$.
Weak annihilation is sensitive to the charge of the decaying $B$ meson and 
thus leads to isospin-breaking differences between $B^+\to\rho^+\gamma$
and $B^0\to\rho^0\gamma$.
The corresponding mechanism is CKM suppressed in the case of $B\to K^*\gamma$,
where penguin operators give the dominant effect for isospin breaking.

\begin{figure}[t]
\center{\resizebox{0.98\textwidth}{!}{%
\includegraphics{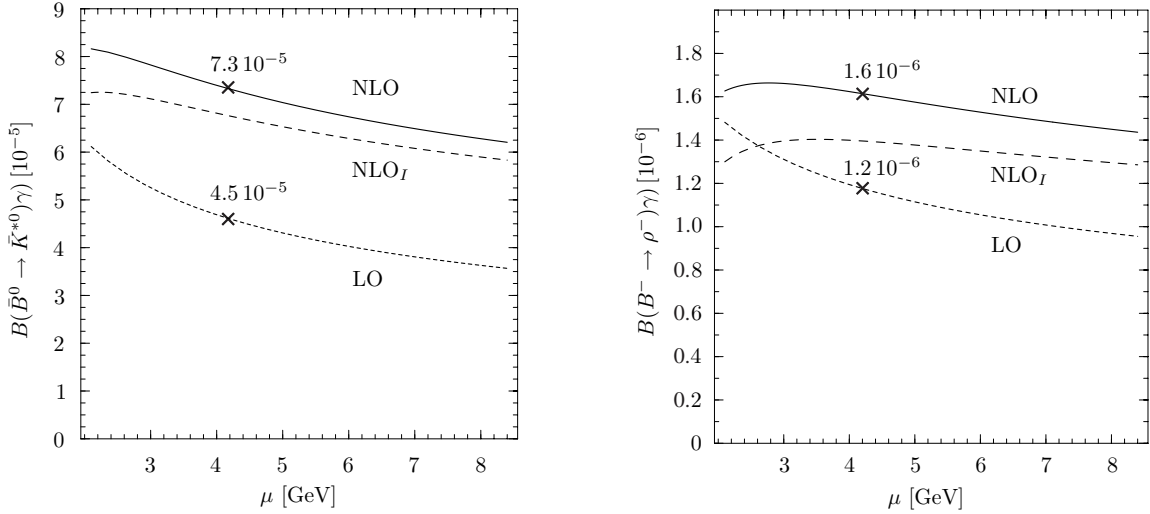}
}}
\caption{Dependence of the branching fractions 
$B(\bar{B}^0 \to \bar{K}^{*0} \gamma)$ and $B(B^- \to \rho^- \gamma)$ on the 
renormalization scale $\mu$. The dotted line shows the LO, the dash-dotted 
line the NLO result including type-I corrections only and the solid line shows 
the complete NLO result.}
\label{fig:bkrhomu}
\end{figure}

The total $\bar B\to V\gamma$ amplitude then can be written as
\begin{equation}
\label{amp}
A(\bar B\to V\gamma)=\frac{G_F}{\sqrt{2}}\left[\lambda_u a^u_7 +
\lambda_c a^c_7\right]\langle V\gamma|Q_7|\bar B\rangle
\end{equation}
where the factorization coefficients $a_7^p(V\gamma)$ consist of the Wilson 
coefficient $C_7$, the contributions from the type-I and type-II 
hard-scattering, and annihilation corrections. One finds a sizeable 
enhancement of the leading order value, dominated by the $T^I$-type 
correction. The net enhancement of $a_7$ at NLO leads to a corresponding 
enhancement of the branching ratios, for fixed value of the form factor. 
This is illustrated in Fig. \ref{fig:bkrhomu},
%%%%%%%%%%%%%%%%%%%%%%%%%%%%%%%%%%%%%%%%%%%%%%%%%%%%%%%%%%%%%%%%%%%
%%%%%%%%%%%%%%%%%%%%%%%%%%%%%%%%%%%%%%%%%%%%%%%%%%%%%%%%%%%%%%%%%%%
where we show the residual scale dependence for 
$B(\bar{B}\to \bar{K}^{*0}\gamma)$ and $B(B^-\to\rho^-\gamma)$ at leading and 
next-to-leading order. 
As shown already in \cite{BBVgam,thesis},
our central values for the $B\to K^{*}\gamma$ branching ratios are 
higher than the experimental measurements (\ref{b0kg}), (\ref{bpkg}). 
The dominant theoretical uncertainty comes from the $B\to V\gamma$ 
form factors. We used the 
light-cone sum rule (LCSR) results $F_{K^{*}}=0.38\pm 0.06$ and 
$F_{\rho}=0.29\pm 0.04$ from \cite{BB98}. 
A recent preliminary lattice QCD determination, 
$F_{K^{*}}=0.25\pm 0.05\pm 0.02$ \cite{SPQCDR}, would give a better 
agreement with the experimental central values.
Further studies of heavy-to-light form factors will also
benefit from developments based on factorization and
soft-collinear effective theory (for recent discussions see 
\cite{PB03,BF03,CLN,BFPS}).

Using the experimental results for the exclusive $B^{0}\to K^{*0}\gamma$ and 
inclusive $B\to X_{s}\gamma$ branching ratios together with their theory 
predictions, we could extract a value of the $B\to K^{*}$ form factor that is 
essentially independent of CKM factors and potential new physics effects. 
The most recent 
measurement of the inclusive branching ratio comes from the Belle 
collaboration, which reports \cite{Bellebsgam}
\begin{equation}
B(B\to X_{s}\gamma)^{\rm exp}_{E_{\gamma >1.8\,\rm{GeV}}}=
(3.59\pm 0.47)\cdot 10^{-4}
\end{equation}
in the photon energy range 1.8 GeV $\le E_{\gamma}\le$ 2.8 GeV. 
The theory predicition from a complete NLO QCD calculation 
is \cite {BCMU}
\begin{equation}
B(B\to X_{s}\gamma)^{\rm th}_{E_{\gamma >1.6\,\rm{GeV}}}=
(3.57\pm 0.30)\cdot 10^{-4}
\end{equation}
for a photon energy cutoff at $E_{0}=1.6$ GeV. Using the 
approximate expression for the integrated branching ratio as a function of 
$E_{0}$ in \cite{GM}, we find that 98.7\% of the events with 
$E_{\gamma}>1.6$ GeV have $E_{\gamma}>1.8$ GeV, i.e.
\begin{equation}
B(B\to X_{s}\gamma)^{\rm th}_{E_{\gamma >1.8\,\rm{GeV}}}=
(3.55\pm 0.30)\cdot 10^{-4}
\end{equation}
Since prediction and measurement are in excellent agreement for the
inclusive branching fractions, we may consider this as a confirmation
of SM short-distance physics in $b\to s\gamma$ transitions.
We can then proceed to directly extract $F_{K^*}$ from the measured
$B(B^0\to K^{*0}\gamma)$.

With our theory prediction for the CP averaged $B(B^{0}\to K^{*0}\gamma)$ 
we get to very good approximation
\begin{equation}
F_{K^{*}}=-0.025^{+0.007}_{-0.016} + 0.150^{+0.010}_{-0.002} 
\sqrt{10^5\, B(B^{0}\to K^{*0}\gamma)^{\rm exp}}
\end{equation}
%\begin{equation}
%F_{K^{*}}=-0.025^{+0.007}_{-0.016} + 47.442^{+3.108}_{-0.682} 
%\sqrt{\frac{B(B^{0}\to K^{*0}\gamma)^{\rm exp} 
%B(B\to X_{s}\gamma)^{\rm th}}{B(B\to X_{s}\gamma)^{\rm exp}}}
%\end{equation}
Here the errors are due to the variation of the renormalization scale 
$m_{b}/2\le\mu\le 2m_{b}$, which is the largest source of theoretical
uncertainty \cite{BBVgam}. 
Using (\ref{b0kg}) and adding errors in quadrature we get
\begin{equation}
F_{K^{*}}^{\rm exp}=0.28\pm0.02
\end{equation}

The input parameters used throughout this paper are collected
in Table \ref{tab:input}.
%%%%%%%%%%%%%%%%%%%%%%%%%%%%%%%%%%%%%%%%%%%%%%%%%%%%
%%%%%%%% table of input parameters   (taken from thesis) %%%%%%%%%%%%%%%
%%%%%%%%%%%%%%%%%%%%%%%%%%%%%%%%%%%%%%%%%%%%%%%%%%%%
\begin{table}[htbp]
\renewcommand{\arraystretch}{1.2}
\begin{center}
\begin{tabular*}{140mm}{@{\extracolsep\fill}|c|c|c|c|c|c|}
\hline\hline
\multicolumn{6}{|c|}{CKM parameters and coupling constants}\\
\hline $V_{us}$ & $V_{cb}$ & $\left|V_{ub}/V_{cb}\right|$ & 
$\Lambda_{\overline{MS}}^{(5)}$ & $\alpha$ & $G_F$\\
\hline 0.22 & 0.041 & $0.09 \pm 0.02$ & $(225 \pm 25)$ MeV & 1/137 & $1.166 
\times 10^{-5} 
\mbox{GeV}^{-2}$\\
\hline
\end{tabular*}

\begin{tabular*}{140mm}{@{\extracolsep\fill}|c|c|c|c|c|}
\hline
\multicolumn{5}{|c|}{Parameters related to the $B$ mesons}\\
\hline $m_B$ & $f_B$ \cite{SMR} & $\lambda_B$ & $\tau_{B^+}$ & $\tau_{B^0}$\\
\hline 5.28 GeV & (200$\pm$30) MeV & $(350 \pm 150)$ MeV & 1.67 ps & 1.54 ps\\
\hline
\end{tabular*}

\begin{tabular*}{140mm}{@{\extracolsep\fill}|c|c|c|c|c|c|}
\hline
\multicolumn{6}{|c|}{Parameters related to the $K^*$ meson \cite{BB98}}\\
\hline $F_{K^*}$ & $f_{K^*}^\perp$ & $m_{K^*}$ & $\alpha^{K^*}_1$ & 
$\alpha^{K^*}_2$ & $f_{K^*}$ \cite{BNPV}\\
\hline $0.38 \pm 0.06$ & 185 MeV & 894 MeV & $0.2\pm 0.2$ & 0.04 & 218 MeV\\
\hline\hline
\multicolumn{6}{|c|}{Parameters related to the $\rho$ meson \cite{BB98}}\\
\hline $F_\rho$ & $f_\rho^\perp$ & $m_\rho$ & $\alpha^\rho_1$  & 
$\alpha^\rho_2$ & $f_\rho$ \cite{BNPV}\\
\hline $0.29 \pm 0.04$ & 160 MeV & 770 MeV & 0 & $0.2\pm 0.2$ & 209 MeV\\
\hline\hline
\multicolumn{6}{|c|}{Parameters related to the $\omega$ meson}\\
\hline $F_\omega$ & $f_\omega^\perp$ & $m_\omega$ & $\alpha^\omega_1$  & 
$\alpha^\omega_2$ & $f_\omega$ \cite{BNPV}\\
\hline $0.29 \pm 0.04$ & 160 MeV & 782 MeV & 0 & $0.2\pm 0.2$ & 187 MeV\\
\hline
\end{tabular*}

\begin{tabular*}{140mm}{@{\extracolsep\fill}|c|c|c|c|}
\hline
\multicolumn{4}{|c|}{Quark and W-boson masses}\\
\hline $m_b(m_b)$ & $m_c(m_b)$ & $m_{t,\mbox{\scriptsize pole}}$ & $M_W$\\
\hline $(4.2 \pm 0.2)$ GeV & $(1.3 \pm 0.2)$ GeV & 174 GeV & 80.4 GeV\\
\hline\hline
\end{tabular*}
\end{center}
\caption[]{Summary of input parameters.\label{tab:input}}
\end{table}
%%%%%%%%%%%%%%%%%%%%%%%%%%%%%%%%%%%%%%%%%%%%%%%%%%%%%%%%%%%%%%%%%

\section{CKM Parameters from $B(B\to\rho\gamma)/B(B\to K^*\gamma)$}
\label{sec:ckmrhok}

Ratios of different $B\to V\gamma$ decay modes can give information 
on parameters in the ($\bar\rho,\bar\eta$) unitarity-triangle plane
with reduced hadronic uncertainties.
The most natural choice is the ratio of the neutral $B^0\to\rho^0\gamma$ and 
$B^0\to K^{*0}\gamma$ branching ratios since annihilation effects in
$B^0\to\rho^0\gamma$ are much reduced in comparison with 
$B^\pm\to\rho^\pm\gamma$. 
On the other hand, these effects can be estimated and the charged mode
can also be used for a similar analysis. 

We define
\begin{equation}
  R(B) = \frac{B(B\to\rho\gamma)}{B(B\to K^{*}\gamma)} \qquad\mbox{and}
\qquad R(\bar B) = \frac{B(\bar B\to\rho\gamma)}{B(\bar B\to K^{*}\gamma)}
\end{equation}
where the $B$ mesons have the quark content $B=(\bar b q)$ and 
$\bar B=(b \bar q)$. We will also consider the CP-averaged ratios 
\begin{equation}
R=\frac{B(B\to\rho\gamma) + B(\bar B\to\rho\gamma)}{
 B(B\to K^*\gamma) + B(\bar B\to\bar K^*\gamma)}
\end{equation}
Omitting the negligible effect of direct CP violation in $B\to K^*\gamma$,
that is assuming $B(B\to K^*\gamma)$ $=$ $B(\bar B\to\bar K^*\gamma)$,
we may write for $R\equiv R_{0},R_{\pm}$
\begin{equation}
  R_{0}=\frac{R(B^{0})+R(\bar B^{0})}{2} \qquad\mbox{and}\qquad 
R_{\pm}=\frac{R(B^{+})+R(B^{-})}{2} 
\end{equation}
The ratio $R(B)$ can be expressed as
\begin{equation}
\label{RB}
R(B) = c_{\rho}^{2} \left| \frac{V_{td}}{V_{ts}}\right|^{2} \xi^{-2} \,r_{m} 
\left| \frac{a_{7}^{c}(\rho\gamma)}{a_{7}^{c}(K^{*}\gamma)} \right|^{2} 
\,\left| 1-\delta a \frac{\bar\rho +i\bar\eta}{1-\bar\rho -i\bar\eta}
\right|^{2}
\end{equation}
Here $c_{\rho}=1/\sqrt{2}$ for $\rho=\rho^{0}$ and $c_{\rho}=1$ for 
$\rho=\rho^{\pm}$,
\begin{equation}
 \xi=\frac{F_{K^{*}}}{F_{\rho}} \qquad\qquad r_{m} = 
\left( \frac{m_{B}^{2} -m_{\rho}^{2}}{m_{B}^{2} -m_{K^{*}}^{2}}\right)^{3} =
1.023
\end{equation}
and
\begin{equation}\label{deltaa}
\delta a  = 
\frac{a_{7}^{u}(\rho\gamma)-a_{7}^{c}(\rho\gamma)}{a_{7}^{c}(\rho\gamma)} 
\end{equation}
The coefficients $a_{7}$ in (\ref{deltaa}) are understood to include the 
annihilation contributions. If annihilation effects are neglected 
$\delta a={\cal O}(\alpha_{s})$. The annihilation terms,
on the other hand, contribute to $\delta a$ only at order $\Lambda_{QCD}/m_b$.
To first approximation weak annihilation is induced by the leading
four-quark operators $Q^u_1$ and $Q^u_2$. It enters the coefficients
$a^u_7(\rho\gamma)$ as an additive term given by \cite{BBVgam,thesis}
\begin{equation}\label{annba}
\begin{array}{ccc}
b_u\, a_1 & \ {\rm for}\ & B^\pm\to\rho^\pm\gamma \\
b_d\, a_2 & \ {\rm for}\ & B^0\to\rho^0\gamma 
\end{array}
\end{equation}
Here $a_{1,2}=C_{1,2}+C_{2,1}/3$ and
\begin{equation}
b_u = \frac{4\pi^2}{3}\frac{f_B f_\rho m_\rho}{F_\rho m_B m_b \lambda_B}
\qquad\qquad b_d =\frac{1}{2}b_u
\end{equation}

In the derivation of (\ref{RB}) we have used the identity
\begin{equation}
\lambda_{c} a_{7}^{c} +\lambda_{u} a_{7}^{u} \equiv -\lambda_{t} a_{7}^{c} 
\left( 1-\frac{\lambda_{u}}{\lambda_{t}} \frac{a_{7}^{u}-a_{7}^{c}}{a_{7}^{c}} 
\right)
\end{equation}
and neglected the second term in the brackets in the case of 
$B\to K^{*}\gamma$ where it amounts to a correction of less than 0.2\% for the 
neutral and less than 1\% for the charged mode.

CP averaging (\ref{RB}) and expanding in $\delta a$ we get
\begin{equation}\label{rrdela}
R = c_{\rho}^{2} \left| \frac{V_{td}}{V_{ts}}\right|^{2} \xi^{-2} \,r_{m} 
\left| \frac{a_{7}^{c}(\rho\gamma)}{a_{7}^{c}(K^{*}\gamma)} \right|^{2} 
\left(1+2\,{\rm Re}\,\delta a
\frac{\bar\eta^{2}-\bar\rho(1-\bar\rho)}{(1-\bar\rho)^{2}+\bar\eta^{2}}\right)
\end{equation}

For the case of the neutral modes ($R_0$), the term proportional to
${\rm Re}\,\delta a$ is a small correction. The numerical value and the
errors from various sources, indicated in brackets, are found to be
\begin{equation}\label{reda0}
{\rm Re}\,\delta a_0=0.002\ ^{-0.023}_{+0.052}\,(\mu)
\ ^{+0.048}_{-0.107}\,(\lambda_B)
\ ^{-0.020}_{+0.025}\,(f_B)\ ^{+0.021}_{-0.022}\,(F_\rho)
\ ^{-0.011}_{+0.015}\,(\alpha^\rho_2)
\end{equation}
The scale $\mu$ has been varied between $m_b/2$ and $2 m_b$ and the
remaining input according to Table \ref{tab:input}.
The central value is very small because of a somewhat accidental
cancellation between the ${\cal O}(\alpha_s)$ effects and the annihilation
corrections in $\delta a$.
Adding in quadrature the positive and negative deviations in (\ref{reda0})
we find
\begin{equation}\label{rda01}
{\rm Re}\,\delta a_0=0.0\pm 0.1
\end{equation}

One may note further that the CKM factor multiplying ${\rm Re}\,\delta a$
in (\ref{rrdela}) is small for the region in the ($\bar\rho,\bar\eta$)
plane allowed by the standard fit of the unitarity triangle. In terms of the
CKM angle $\gamma$ and $R_b=\sqrt{\bar\rho^2 +\bar\eta^2}$ this factor
can be written as
\begin{equation}\label{frhoeta}
f(\bar\rho,\bar\eta)\equiv 
\frac{\bar\eta^{2}-\bar\rho(1-\bar\rho)}{(1-\bar\rho)^{2}+\bar\eta^{2}}=
\frac{R^2_b-R_b \cos\gamma}{1-2 R_b \cos\gamma + R^2_b}
\end{equation}
The standard fit region, which is the most interesting for precision
tests of the CKM framework, is roughly characterized by
\begin{equation}
0.3\leq R_b \leq 0.5\qquad\qquad 
\frac{\pi}{4} \leq\gamma\leq \frac{\pi}{2}
\end{equation}
This implies $-0.2\leq f(\bar\rho,\bar\eta)\leq 0.2$. Together with
(\ref{rda01}) we then have
\begin{equation}\label{rda0f}
|{\rm Re}\,\delta a_0\,f(\bar\rho,\bar\eta)| < 0.02
\end{equation}
This means that, under the conditions mentioned above, the
correction proportional to ${\rm Re}\,\delta a$ in (\ref{rrdela}) can be
safely neglected and the relation between $R_0$ and CKM quantities
greatly simplifies.

Taking into account the uncertainties from scale dependence, $\lambda_B$,
$f_B$, $F_{K^*}$, $F_\rho$, $\alpha^{K^*}_1$ and $\alpha^\rho_2$, we get
\begin{equation}
\left|\frac{a^c_7(\rho\gamma)}{a^c_7(K^*\gamma)}\right| =1.01\pm 0.02
\end{equation}
We recall that $a^c_7$ is essentially free of annihilation contributions,
which mainly affect $a^u_7$. Defining
\begin{equation}
\kappa^{-1}\equiv\sqrt{r_m}\,
\left|\frac{a^c_7(\rho\gamma)}{a^c_7(K^*\gamma)}\right|\qquad\qquad
\kappa=0.98\pm 0.02
\end{equation}
and recalling (\ref{Rtdef}), we finally have
\begin{equation}\label{rtr0}
R_t=\sqrt{2}\frac{\kappa}{\lambda}\,\xi\,\sqrt{R_0}
=0.82\frac{\xi}{1.3}\sqrt{\frac{R_0}{0.01}}
\end{equation}  
Using $\kappa=0.98$, which leads to the second equality in (\ref{rtr0}),
this formula holds to within $\pm 3\%$. In this approximation
$R_t$, which is the radius of a circle around the 
point $(1,0)$ in the $(\bar\rho,\bar\eta)$ plane, is directly given
in terms of the CP-averaged ratio of branching fractions
$R_0=B(B^0\to\rho^0\gamma)/B(B^0\to K^{*0}\gamma)$.
The theoretical uncertainty is essentially reduced to the
SU(3) breaking parameter $\xi=F_{K^*}/F_\rho$. We use the LCSR estimate 
$\xi =1.31\pm 0.13$ \cite{BB98}. 
A preliminary lattice value is $\xi =1.1\pm 0.1$ \cite{SPQCDR}.

In the case of the charged modes, with a decaying $B^\pm$, weak
annihilation dominates $\delta a$ and we typically have
\begin{equation}
{\rm Re}\,\delta a_\pm = -0.4\pm 0.4
\end{equation}
The uncertainty is largely due to $\lambda_B$ determining the strength
of weak annihilation. This parameter is still not well known at present, 
but the situation can in principle be systematically improved
\cite{BK,BIK}.

\begin{figure}[t]
\center{\resizebox{0.8\textwidth}{!}{%
\includegraphics{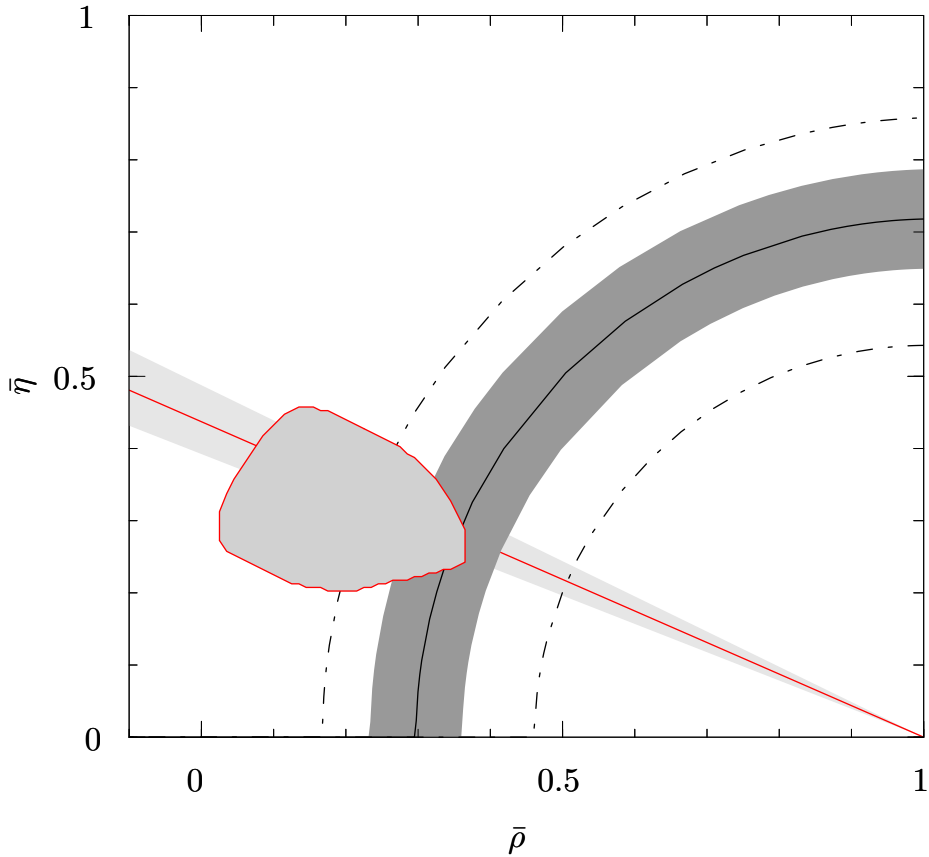}
}}
\caption{Constraints implied by $R_0$ in the ($\bar\rho,\bar\eta$) plane. 
The experimental value used is $R_0=0.007\pm 0.003$
(see text for further explanation). The width of the dark band reflects a 
$\pm 10\%$ variation of $\xi$ for central $R_0$. 
The dash-dotted lines display the error from
$R_0$ while $\xi$ is kept at its central value. 
%The dashed curve
%represents a lower limit on $R_t$ based on the assumption
%$R_0 > 0.0146$ and $\xi > 1$. 
The region obtained from a standard fit of the
unitarity triangle (irregularly shaped area) and the constraint from
$\sin 2\beta$ (light shaded band) are overlaid.}
\label{fig:R00}
\end{figure}

The constraint in the ($\bar\rho,\bar\eta$) plane implied by a measurement
of $R_0$ is shown in Fig. \ref{fig:R00}. 
%%%%%%%%%%%%%%%%%%%%%%%%%%%%%%%%%%%%%%%%%%%%%%%%%%%%%%%%%%%%%%%%%%%
% export Mathematica plot as eps file and use epstopdf to get the 
% bounding box correctly
%%%%%%%%%%%%%%%%%%%%%%%%%%%%%%%%%%%%%%%%%%%%%%%%%%%%%%%%%%%%%%%%%%%
For the purpose of illustration we shall assume that the results
in \cite{Brhoexp1} and \cite{Brhoexp2} can be interpreted to give
\begin{equation}\label{br0gbelle}
B(B^0\to\rho^0\gamma)=\left(0.30 \pm 0.12\right)\cdot 10^{-6}
\end{equation}
Here we have combined the average $\rho^\pm/\rho^0/\omega$
branching ratio from \cite{Brhoexp1} and \cite{Brhoexp2}, obtaining
$B(B\to (\rho/\omega)\gamma)=(0.64\pm 0.27)\cdot 10^{-6}$.
Dividing by $2\,\tau_{B^+}/\tau_{B^0}$ then gives  (\ref{br0gbelle})
as an estimate for $B(B^0\to\rho^0\gamma)$.
We use the central value in (\ref{b0kg}) to compute
the experimental ratio 
\begin{equation}\label{r0num}
R_0=\frac{B(B^0\to\rho^0\gamma)}{B(B^0\to K^{*0}\gamma)}
=0.007\pm 0.003
\end{equation}
Adopting an error of $\pm 10\%$ for the SU(3)-breaking
form factor ratio $\xi=1.31\pm 0.13$ and the central value in
(\ref{r0num}), we obtain the dark shaded band 
in Fig. \ref{fig:R00}. Here the full expression (\ref{RB}) is used, without
expanding in $\delta a$, and all theoretical parameters besides $\xi$
are kept at their central values. This is justified as the theoretical
uncertainty is entirely dominated by $\xi$.
For the same constraint, the dash-dotted lines indicate the $1\sigma$ 
experimental uncertainty from (\ref{r0num}) with fixed $\xi=1.3$.

As can be seen, the intersection of the constraints from
$R_0$ and $\sin 2\beta$ determines the apex ($\bar\rho,\bar\eta$)
of the unitarity triangle.
For comparison, the standard fit region for the unitarity triangle
in the ($\bar\rho,\bar\eta$) plane \cite{ckmfitter}
and the constraint from the experimental measurement of 
$\sin 2\beta=0.734\pm 0.054$ \cite{sin2b} are also shown in 
Fig. \ref{fig:R00}.

\begin{figure}[t]
\center{\resizebox{0.8\textwidth}{!}{%
\includegraphics{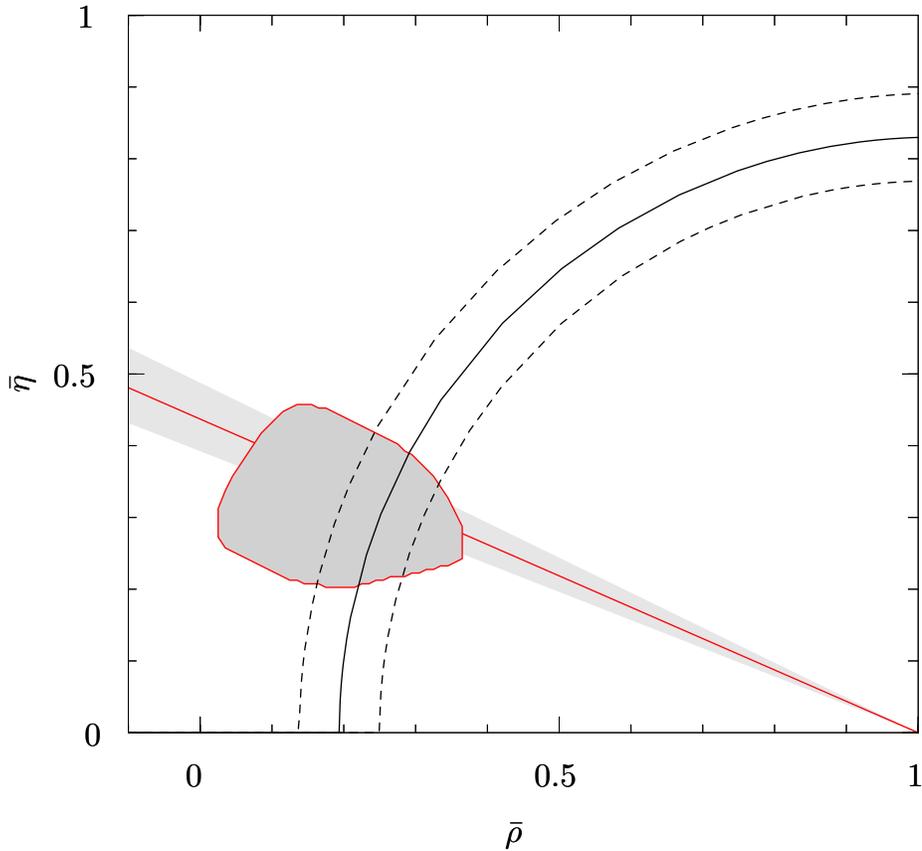}
}}
\caption{Upper bound on $R_t$ (the distance from point $(1,0)$)
implied by the Babar limit $B(B^0\to\rho^0\gamma) < 0.4\cdot 10^{-6}$ 
in the ($\bar\rho,\bar\eta$) plane. 
The curves correspond (from left to right) to 
$\xi\equiv F_{K^*}/F_\rho =1.4$, $1.3$ and $1.2$.
The region obtained from a standard fit of the
unitarity triangle (irregularly shaped area) and the constraint from
$\sin 2\beta$ (light shaded band) are overlaid.}
\label{fig:R00UL}
\end{figure}

We finally note that the information from $R_0$ 
is already becoming comparable with the 
constraint from the ratio of $B_{d}$ and $B_{s}$ meson mixing
frequencies $\Delta M_{B_{d}}$ and $\Delta M_{B_{s}}$ \cite{deltaBs}. 
It is possible that a useful experimental measurement of $R_{0}$ might 
actually be achieved before the measurement of $\Delta M_{B_{s}}$. 
Very interesting in this respect is the recent
upper bound for $B(B^0\to\rho^0\gamma)$  from Babar (\ref{b0rg}).
As we have discussed above, the neutral mode is favoured theoretically
because of the small impact of annihilation effects.
In addition, it turns out that the upper bound for $B(B^0\to\rho^0\gamma)$
is particularly strong in comparison with the bound for the
charged mode (\ref{bprg}), even after correcting for an isospin factor
of 2 and the $B^+/B^0$ lifetime difference. 
We thus prefer to use the neutral mode directly for placing an upper bound
on $R_t$, rather than the combined result of the three modes 
$(\rho^\pm/\rho^0/\omega)\gamma$, which was the choice made in 
\cite{Brhoexp1}.
Using (\ref{rtr0}), the recent Babar limit (\ref{b0rg}) together with
(\ref{b0kg}) implies
\begin{equation}\label{rtlim}
R_t < 0.82 \,\frac{\xi}{1.3}
\end{equation}
This is equivalent to
\begin{equation}\label{vtdlim}
\left|\frac{V_{td}}{V_{ts}}\right| < 0.18 \,\frac{\xi}{1.3} \qquad\qquad
|V_{td}| < 7.3\cdot 10^{-3} \,\frac{\xi}{1.3} 
\end{equation}
The bound may be compared with the
$2\sigma$ range 
\begin{equation}\label{vtdran}
6.5\cdot 10^{-3} <  |V_{td}| < 9.5 \cdot 10^{-3}  
\end{equation}
obtained from a standard fit of the unitarity
triangle \cite{ckmfitter}.
For $30\%$ SU(3) breaking in the ratio of form factors, $\xi =1.3$,
more than half of the range (\ref{vtdran}) is excluded by (\ref{vtdlim}).
Should the amount of SU(3) breaking be less than $30\%$, the bound
would be even stronger.
An illustration of the Babar bound in the $(\bar\rho,\bar\eta)$ plane
is given in Fig. \ref{fig:R00UL}.
%%%%%%%%%%%%%%%%%%%%%%%%%%%%%%%%%%%%%%%%%%%%%%%%%%%%%%%%%%%%%%%%%%%
% export Mathematica plot as eps file and use epstopdf to get the 
% bounding box correctly
%%%%%%%%%%%%%%%%%%%%%%%%%%%%%%%%%%%%%%%%%%%%%%%%%%%%%%%%%%%%%%%%%%%

\section{$B\to\rho\gamma$ and $B\to\rho l\nu$}

As we have seen, the rare decay $B\to\rho\gamma$ is a clean probe
of flavour physics, except for the sizable uncertainty in the form
factor $F_\rho$. Other uncertainties are quite well under control
within a treatment of the decay at next-to-leading order in QCD
and a leading-order evaluation of power-suppressed annihilation effects.
This is the case in particular for the neutral channel 
$B^0\to\rho^0\gamma$, where weak-annihilation effects are small.
The sensitivity to $F_\rho$ can be reduced by taking the ratio of
$B\to\rho\gamma$ and $B\to K^*\gamma$ branching fractions, as we
have discussed in the previous section. Then the impact of long-distance
hadronic physics is limited to SU(3) breaking in the ratio 
$\xi=F_{K^*}/F_\rho$. While this is certainly an advantage,
the exact deviation from the SU(3) limit
$\xi=1$ remains at present a significant source of uncertainty.

In this section we discuss a possibility to reduce hadronic
uncertainties in a different way, using the ratio of $B\to\rho l\nu$
and $B\to\rho\gamma$ decay rates. The simplification occurs because relations 
exist between the corresponding form factors in the large energy limit.
Since only $B\to\rho$ transitions are involved, the problems with SU(3)
breaking are avoided and only isospin symmetry needs to be assumed,
which should be valid to within a few percent. The existence of
relations between the form factors in the large energy limit and their 
potential usefulness for phenomenology were first pointed out in 
\cite{CLOPR}. The results of \cite{CLOPR} were put on a field theoretical
basis within the soft-collinear effective theory (SCET) and extended to
higher order in QCD \cite{BF,BFPS,CLN}. These relations were
applied to extract information on the form factor in $B\to K^*\gamma$
for use in other channels such as $B\to K^*l^+l^-$ \cite{BH}.
Previously, the authors of \cite{BD} have investigated the possibility
to relate $B\to\rho l\nu$ in a certain region of phase space with
$B\to K^*\gamma$. This suggestion is similar in spirit to our proposal,
but the analysis of \cite{BD} was based only on the heavy quark limit,
instead of the full large energy relations from \cite{CLOPR}, and
was still affected by SU(3) breaking.
In addition, our discussion also includes short-distance QCD corrections
at next-to-leading order.

The differential decay rate for $B\to\rho l\nu$ is given by
\begin{equation}\label{gbrln}
\frac{d^2\Gamma(B\to\rho l\nu)}{ds\, dz}=c^2_\rho
\frac{G^2_F m^5_B}{256 \pi^3}|V_{ub}|^2 w^{1/2}
\left[ \frac{(1-z)^2}{2} H^2_+ + \frac{(1+z)^2}{2} H^2_- +(1-z^2) H^2_0
\right]
\end{equation}
Here
\begin{equation}
w\equiv w(s,r)=1+s^2 + r^2 -2 s-2r-2 s r
\end{equation}
and
\begin{equation}
z=\cos\theta \qquad\qquad s=\frac{q^2}{m^2_B}\qquad\qquad 
r=\frac{m^2_\rho}{m^2_B}\approx 0.021 
\end{equation}
where $q^2$ is the dilepton invariant mass and $\theta$ is the angle
between the momenta of the neutrino and the $B$ meson in the
dilepton centre-of-mass frame. Equivalently, $\theta$ is the angle
between the charged-lepton momentum and the direction anti-parallel to the 
$B$ momentum in this frame. The same definition of $\theta$ is valid
for $B\to\rho l\nu$ with either a positive or a negative charged lepton.
The kinematical range for $s$ and $z$ is
\begin{equation}
0\leq s\leq (1-\sqrt{r})^2\qquad -1\leq z\leq 1
\end{equation}
The $H_i\equiv H_i(s)$ are helicity form factors. They can be 
expressed in terms of the vector and axial vector form factors
$V(s)$, $A_1(s)$ and $A_2(s)$ as
\begin{eqnarray}
H_\pm &=& 
\sqrt{s}\left[(1+\sqrt{r})A_1\mp\frac{\sqrt{w}}{1+\sqrt{r}}\, V\right]
\label{hpm}\\
H_0 &=& \frac{1}{2\sqrt{r}}
\left[(1-s-r)(1+\sqrt{r})A_1 - \frac{w}{1+\sqrt{r}} A_2\right]\label{h0}
\end{eqnarray}
where we use the conventions of \cite{CLOPR} for $V$, $A_1$, $A_2$,
which, in particular, are positive real quantities.

The CP-averaged decay rate for $B\to\rho\gamma$ can be
written as
\begin{equation}\label{gbrgcpa}
\Gamma(B\to\rho\gamma)=\frac{G^2_F\alpha m^3_B m^2_b}{32\pi^4}
  (1-r)^3 \,|V_{td} V_{tb}|^2 \,|a^c_7(\rho\gamma)|^2\, c^2_\rho F^2_\rho
\end{equation}
where we have used the approximation, explained in the previous
section, that corresponds to neglecting the term $\sim {\rm Re}\,\delta a$
in (\ref{rrdela}). As we have seen, this is a very good approximation 
for $B^0\to\rho^0\gamma$. If a more accurate treatment is desired, or the
analysis should be applied to $B^\pm\to\rho^\pm\gamma$, the
following discussion can be generalized in a straightforward way using
the complete expression based on (\ref{amp}).
Combining (\ref{gbrln}) and (\ref{gbrgcpa}) we find for the case of
neutral $B$ mesons
\begin{equation}\label{brlnbrg}
k(s,z)
\left|\frac{V_{ud}V_{ub}}{V_{td}V_{tb}}\right|^2 =
\frac{4\alpha}{\pi}\,\frac{m^2_b}{m^2_B}
\,\frac{(1-r)^3 |V_{ud}|^2 |a^c_7(\rho\gamma)|^2}{B(B^0\to\rho^0\gamma)}
\ \frac{dB(B^0\to\rho l\nu)}{ds\, dz}
\end{equation}
Here $B^0\to\rho l\nu$ can be either one of the two
channels $B^0\to\rho^+ l^-\bar\nu$ or $B^0\to\rho^- l^+\nu$
and $l$ may be an electron or a muon.
The hadronic quantity $k$ is defined as
\begin{equation}\label{kdef}
k(s,z)=w(s,r)^{1/2}\left[ \frac{(1-z)^2}{2} \frac{H^2_+}{F^2_\rho} + 
 \frac{(1+z)^2}{2} \frac{H^2_-}{F^2_\rho} +(1-z^2) \frac{H^2_0}{F^2_\rho}
\right]
\end{equation}
The differential branching ratio and the function $k$ in (\ref{brlnbrg})
may be replaced by their integrated versions
\begin{equation}\label{bkint}
\Delta B(\bar s,\epsilon)=
\int_0^{\bar s}ds\int_{1-\epsilon}^1 dz \frac{dB(B^0\to\rho l\nu)}{ds\, dz}
\qquad
K(\bar s,\epsilon)= \int_0^{\bar s}ds\int_{1-\epsilon}^1 dz\, k(s,z)
\end{equation}
where
\begin{equation}\label{sbeps}
0\leq \bar s \leq s_{max}\equiv (1-\sqrt{r})^2 \approx 0.73
\qquad\quad
0\leq \epsilon\leq 2
\end{equation}
such that the fully integrated branching fraction for $B^0\to\rho l\nu$ 
is given by $\Delta B(s_{max},2)$.

The relation (\ref{brlnbrg}) allows us to determine the
CKM parameter
\begin{equation}\label{vubvtd}
\left|\frac{V_{ud}V_{ub}}{V_{td}V_{tb}}\right|^2 =
\frac{\bar\rho^2 +\bar\eta^2}{(1-\bar\rho)^2+\bar\eta^2}
\end{equation}
in terms of observable $B\to\rho\gamma$ and $B\to\rho l\nu$
branching fractions, known quantities, and the hadronic function $k$.
The main virtue of this expression is that in the large energy limit
hadronic form factors cancel in the ratios $H_\pm/F_\rho$. This is not
the case for $H_0/F_\rho$, but its contribution can be suppressed
by selecting events in the vicinity of $z=1$.
As a consequence, (\ref{brlnbrg}) can be turned into a
theoretically clean expression for the determination of the CKM ratio
in (\ref{vubvtd}).

In the large energy limit the form factors $H_\pm$, $H_0$ and $F_\rho$
can be written in terms of just two independent form factors
$\zeta_\perp(s)$ and $\zeta_\parallel(s)$ using \cite{CLOPR}
\begin{equation}\label{a1v}
A_1(s) =\frac{1-s+r}{1+\sqrt{r}}\zeta_\perp(s)
\qquad\quad V(s)=(1+\sqrt{r})\,\zeta_\perp(s)
\end{equation}
\begin{equation}\label{a2f}
A_2(s) =(1+\sqrt{r})\left[\zeta_\perp(s)-
\frac{2\sqrt{r}}{1-s+r}\zeta_\parallel(s)\right]
\qquad\quad F_\rho\equiv T_1(0)=\zeta_\perp(0)
\end{equation}
Together with (\ref{hpm}), (\ref{h0}) these relations imply
\begin{eqnarray}
H_\pm(s) &=& \sqrt{s}\left[\,1-s+r\mp\sqrt{w}\,\right]\,\zeta_\perp(s)
\label{hpmz}\\
H_0(s) &=& \sqrt{r}(1+s-r)\,\zeta_\perp(s) +\frac{w}{1-s+r}\,\zeta_\parallel(s)
\label{h0z}
\end{eqnarray}
These results are valid in the heavy-quark limit and the limit
of large energy of the recoiling $\rho$-meson
\begin{equation}
E_\rho =\frac{m_B}{2}(1-s+r)
\end{equation}
In this approximation the ratio $H_+/H_-$ is independent of
hadronic form factors. For not too large values of $s$ this ratio
is strongly suppressed, $H_+/H_-={\cal O}(r)$.
More importantly, also $H_-(s)$ and $F_\rho$ depend on the same
form factor $\zeta_\perp(s)$, which is to be evaluated at $s=0$
in the latter case. As a consequence, we may write
\begin{equation}\label{hmfrho}
\frac{H^2_-(s)}{F^2_\rho}=4s\, (1+ b_1 s + \ldots)
\end{equation}
expanding the form factor ratio in a Taylor series.
The leading term in this ratio for small $s$ is largely free of hadronic
uncertainties in the large energy limit. The higher-order
corrections only depend on the shape of $\zeta_\perp(s)$, not on its
absolute normalization, and can in principle be determined from a
fit to the shape of the observed spectrum in $s$.
The coefficient $b_1$ is related to the slope of $\zeta_\perp$
and can be written as
\begin{equation}\label{b1zeta}
b_1=2\left(\frac{\zeta'_\perp(0)}{\zeta_\perp(0)}-\frac{1}{1-r}\right)
\end{equation}
When fitting the ratio in (\ref{hmfrho}) to the experimental
spectrum, other parametrizations for the shape may, of course, 
be chosen. The Taylor series $1+b_1 s+\ldots$ could be replaced
for instance by the pole form $1/(1-\beta_1 s)^2$, or a combination
of the two.

In \cite{BF} the corrections of order $\alpha_s$ have been computed
to the relations between form factors in the large energy limit.
There is no relative correction between $A_1$ and $V$ to all
orders in $\alpha_s$ \cite{HBLN}.
Therefore the correction of the ratio $V/F_\rho$ given in \cite{BF}
also applies to $H_-/F_\rho$.
Taking these effects into account, the leading term ($4s$) in
(\ref{hmfrho}) is modified to
\begin{equation}\label{hmfals}
\frac{H^2_-(s)}{F^2_\rho}=4s\,\left(1+\frac{2\alpha_s(\mu_1)}{3\pi}
\left[1+2 \ln\frac{\mu_1}{m_b}\right]
-\frac{\alpha_s(\mu_2)}{3\pi}\frac{\Delta F_\perp}{V(0)}(1+\sqrt{r})
\right)
\end{equation}
where the first term with $\alpha_s(\mu_1)\approx 0.22$ refers
to the vertex correction and the second with $\alpha_s(\mu_2)\approx 0.34$
to the hard spectator interaction. 
The usual renormalization scheme of the form factor $F_\rho$, 
adopted in this paper and used in (\ref{hmfals}), 
corresponds to the $\overline{MS}$ scheme with anticommuting $\gamma_5$ (NDR). 
Numerically, the QCD correction factor amounts to $(1-(0.15\pm 0.10))$
using the estimates in \cite{BF}. The dominant uncertainty comes from
$\Delta F_\perp$, which depends on properties of the $B$-meson
light-cone wave function. This quantity is poorly known at present,
but improvements should be possible in the future and would lead to
a reduction in the uncertainty.

It is interesting to compare the above analysis with the results
for the form factors obtained using the method of light-cone QCD sum
rules \cite{BB98}. With the form factors computed in \cite{BB98}
one finds
\begin{equation}\label{hmflcsr}
\frac{H^2_-(s)}{F^2_\rho}=4.25 s\, (1+ 0.59 s + 0.65 s^2 + \ldots)
\end{equation}
The leading term agrees very well with the prediction at leading
order in the large energy limit (\ref{hmfrho}). Taking the QCD corrections
into account according to (\ref{hmfals}), the prediction for this term
is typically about $15\%$ lower. On the other hand, the result in
(\ref{hmflcsr}) has an uncertainty of about $30\%$ \cite{BB98}.
Nevertheless, the general level of agreement of the sum rule calculations,
which include subleading corrections in $1/m_b$, with the large energy limit,
is consistent with the assumption that power corrections are of
moderate size.

In contrast to $H_\pm$, the longitudinal form factor $H_0$ is dominated
by $\zeta_\parallel$, which is not cancelled in the ratio
$H_0/F_\rho$. The third term in (\ref{kdef}) can still be estimated
theoretically, but will be affected by larger uncertainites.
As mentioned above, in order to reduce its importance a cut on
the angular variable $z$ may be imposed, restricting $z$ to
be in the vicinity of $+1$ or $-1$. The latter case is not interesting,
since it would strongly suppress the $H_-$ contribution, leaving
only the contribution from $H_+$, which is very small.
Parametrizing the cut below $z=1$ by $\epsilon$ as defined in (\ref{bkint})
and performing the angular integration, we find for $K(\bar s,\epsilon)$
\begin{equation}\label{kinte}
K(\bar s,\epsilon)=\int^{\bar s}_0 ds\, \sqrt{w}\left[
\frac{\epsilon^3}{6}\frac{H^2_+}{F^2_\rho}+
\left(2\epsilon-\epsilon^2+\frac{\epsilon^3}{6}\right)\frac{H^2_-}{F^2_\rho}+
\left(\epsilon^2-\frac{\epsilon^3}{3}\right)\frac{H^2_0}{F^2_\rho}
\right]
\end{equation}
The full angular range is obtained for $\epsilon=2$ and in this case
all three $\epsilon$-dependent coefficients become equal to $4/3$.
For small $\epsilon$, on the other hand, a strong hierarchy exists,
which is clearly visible in (\ref{kinte}). The contribution from
$H_0$ is suppressed with respect to the $H_-$-term by a factor
of $\epsilon/2$, that is by one order of magnitude for $\epsilon=0.2$.
The corresponding suppression of the $H_+$-term is even by a factor
of $\epsilon^2/12$, in addition to the fact that $H_+/H_-$ is already small
for moderate values of $s$. The contribution from $H_+$ is therefore
entirely negligible in the following discussion.

Neglecting all terms of ${\cal O}(\epsilon^3)$, (\ref{kinte})
simplifies to
\begin{equation}\label{kinte2}
K(\bar s,\epsilon)=\int^{\bar s}_0 ds\, \sqrt{w}\left[
2\epsilon\, \frac{H^2_-}{F^2_\rho}+
\epsilon^2\frac{H^2_0-H^2_-}{F^2_\rho}\right]
\end{equation}
The validity of the large energy limit, with the model-independent
normalization of $H^2_-/F^2_\rho$ in (\ref{hmfrho}), (\ref{hmfals}),
requires moderate values of $s$. Enhancing this term in (\ref{kinte})
requires small $\epsilon$.
As a typical example one may concentrate on the part of phase space
defined by $\bar s=0.4$ and $\epsilon=0.2$.
The relative number of $B^0\to\rho l\nu$ events in this region
($0\leq s\leq 0.4$, $0.8\leq z\leq 1$) is given by
\begin{equation}\label{k0402}
\frac{K(0.4,0.2)}{K(s_{max},2)}\approx 0.064
\end{equation}
For this estimate we have evaluated $K(\bar s,\epsilon)$  
employing the form factors from \cite{BB98}.
A measurement of $B\to\rho l\nu$ has been reported by CLEO \cite{CLEO},
\begin{equation}\label{bbrlnexp1}
B(B^0\to\rho l\nu)=(2.17\pm 0.73)\cdot 10^{-4}
\end{equation}
and BaBar \cite{Bulnuexp}:
\begin{equation}\label{bbrlnexp2}
B(B^0\to\rho l\nu)=(2.57\pm 0.79)\cdot 10^{-4}
\end{equation}
The effective branching ratio of $B\to\rho l\nu$ events in the
above region of phase space would then be about $10^{-5}$.

\begin{figure}[t]
\center{\resizebox{0.5\textwidth}{!}{%
\includegraphics{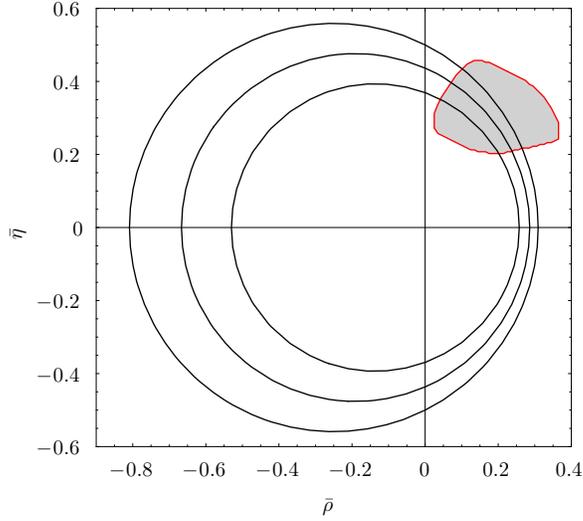}
}}
\caption{Constraints in the ($\bar\rho,\bar\eta$) plane
implied by $|V_{ud}V_{ub}/V_{td}|^2 =0.16\pm 0.04$.
For comparison, the standard fit region is indicated by
the shaded area. 
}
\label{fig:vubvtd}
\end{figure}

The first term in (\ref{kinte2}) is determined by the measured shape
of the $s$-distribution and the model-independent normalization
in (\ref{hmfrho}), (\ref{hmfals}). The small correction from
the second term in (\ref{kinte2}) could either be estimated theoretically,
or be isolated in the data by varying $\epsilon$.
With the form factors from \cite{BB98} we have for instance
\begin{equation}
K(0.4,\epsilon)=0.57\,\epsilon + 0.25\,\epsilon^2
\end{equation}
Once $K(\bar s,\epsilon)$ is known, the measured values of
$\Delta B(\bar s,\epsilon)$ (\ref{bkint}) and $B(B^0\to\rho^0\gamma)$
determine the CKM quantity in (\ref{vubvtd}) using (\ref{brlnbrg}).
This CKM ratio provides us with an interesting constraint in the
($\bar\rho,\bar\eta$) plane, which is illustrated in Fig. \ref{fig:vubvtd}
for a hypothetical measurement 
of $|V_{ud}V_{ub}/V_{td}|^2 =0.16\pm 0.04$.
%%%%%%%%%%%%%%%%%%%%%%%%%%%%%%%%%%%%%%%%%%%%%%%%%%%%%%%%%%%%%%%%%%%
%%%%%%%%%%%%%%%%%%%%%%%%%%%%%%%%%%%%%%%%%%%%%%%%%%%%%%%%%%%%%%%%%%%
We observe that the constraint is quite stringent, in particular
in the important region corresponding to the standard fit results,
and even for the rather moderate precision of $\pm 25\%$.

\section{Isospin Breaking in $B\to V\gamma$}
\label{sec:ibrhogam}

The CP averaged isospin breaking ratio can be defined as
\begin{equation}
\Delta (V\gamma)=\frac{\Gamma (B^{0}\to V^{0}\gamma) - 
v \Gamma(B^{\pm}\to V^{\pm}\gamma)}{\Gamma (B^{0}\to V^{0}\gamma) + 
v \Gamma(B^{\pm}\to V^{\pm}\gamma)}
\end{equation}
with $v=1$ for $V=K^{*}$ and $v=1/2$ for $V=\rho$. This ratio has a reduced 
sen\-si\-ti\-vi\-ty to the nonperturbative form factors. 
As already discussed, in our approximations, isospin breaking is generated by 
weak annihilation contributions. Kagan and Neubert found a large effect 
from the penguin operator $Q_{6}$ on the isospin asymmetry 
$\Delta (K^{*}\gamma)$ \cite{KN}. 
Our prediction 
$\Delta(K^*\gamma)=(3.9^{+3.1}_{-1.9})\%$ (see \cite{thesis}) 
is in agreement with the experimental results (Belle in \cite{BKgamexp},
Babar \cite{EP})
\begin{eqnarray}
\Delta(K^*\gamma) &=& +0.034 \pm 0.044 \pm 0.026 \pm 0.025 
\ \ \ \ {\rm (Belle)}\\
\Delta(K^*\gamma) &=& +0.051 \pm 0.044 \pm 0.023 \pm 0.024 
\ \ \ \ {\rm (Babar)}
\end{eqnarray} 
Here the errors are statistical, systematic and from the $B^+/B^0$ production
ratio.

%$\Delta(K^*\gamma)_{exp}=(4.3\pm 4.8)\%$ from (\ref{b0kg}), (\ref{bpkg}). 

\begin{figure}[t]
\center{\resizebox{0.98\textwidth}{!}{%
\includegraphics{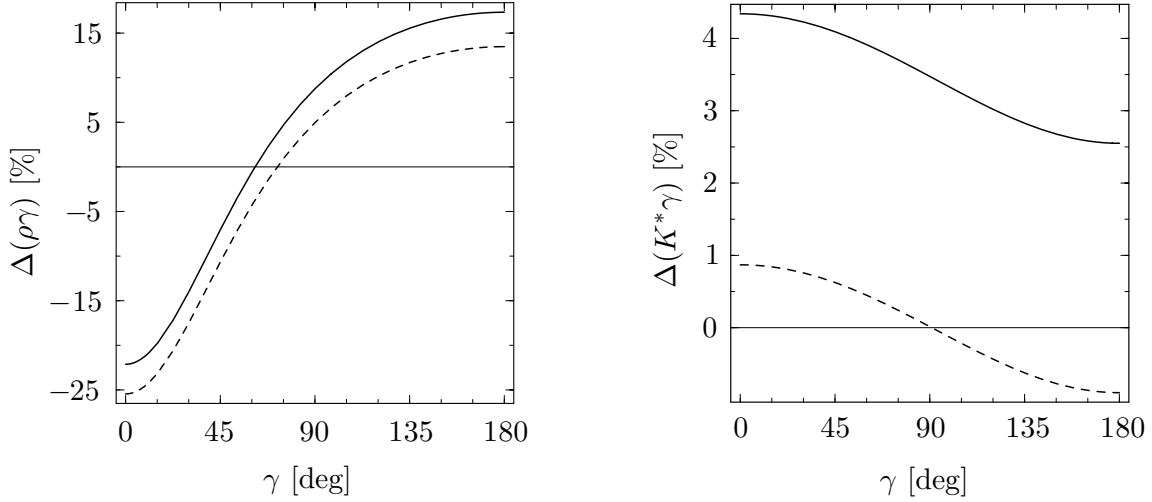}
}}
%\vspace{-1.7cm}
\caption[1]{The isospin-breaking asymmetries $\Delta(\rho\gamma)$ and
$\Delta(K^*\gamma)$ as a function of the CKM angle $\gamma$ with (solid)
and without (dashed) the inclusion of QCD penguin operator effects.}
\label{fig:isodelta}
\end{figure}

For $B\to\rho\gamma$ we find a 
strong dependence of the isospin asymmetry on the angle $\gamma$ of the 
unitarity triangle. As seen in Fig.~\ref{fig:isodelta},
%%%%%%%%%%%%%%%%%%%%%%%%%%%%%%%%%%%%%%%%%%%%%%%%%%%%%%%%%%%%%%%%%%%
% figure created very complicatedly (because Mathematica creates same 
%bounding box for both plots and not same plot frame box): created 
%Mathematica plots, used lllustrator to add labels, make plots same size 
%and save as .pdf and .eps files
%%%%%%%%%%%%%%%%%%%%%%%%%%%%%%%%%%%%%%%%%%%%%%%%%%%%%%%%%%%%%%%%%%%
the $\gamma$ dependence is in particular pronounced for the zero crossing of 
$\Delta (\rho\gamma)$ around $\gamma=60^{\circ}$, the value favoured by the 
standard UT fits. 

\begin{figure}[t]
\center{\resizebox{0.8\textwidth}{!}{%
\includegraphics{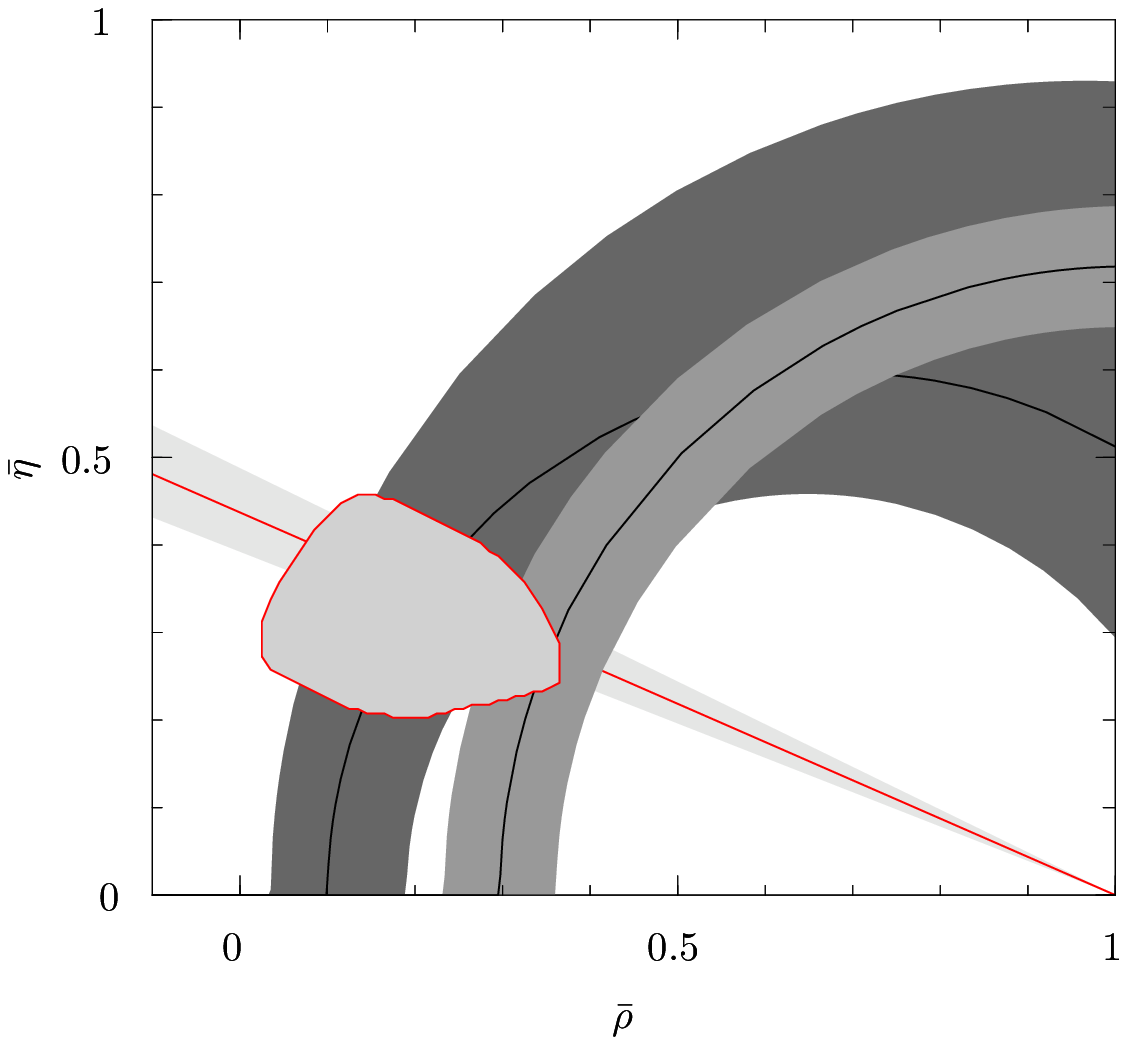}
}}
\caption{Same as Fig.~\ref{fig:R00} including the implication of a 
measurement of $\Delta (\rho\gamma)_{exp}=0$ (curved band 
on the right). The width of the band reflects the theoretical uncertainties 
from varying the hadronic parameter $\lambda_{B}$ and the renormalization 
scale $\mu$. (The effect of isospin breaking in the form factors is neglected 
here.)}
\label{fig:R00iso}
\end{figure}

Once a measurement of both the charged and neutral $B\to\rho\gamma$ modes is 
available, the isospin-asymmetry $\Delta(\rho\gamma)$ can be used to 
constrain the unitarity triangle. For the purpose of illustration we plot in 
Fig.~\ref{fig:R00iso}, in addition to the $R_0$ and $\sin 2\beta$ bands
shown already in Fig. \ref{fig:R00}, the implication of an assumed 
measurement of 
$\Delta (\rho\gamma)_{exp}=0$, which would correspond to 
the Standard Model prediction for a CKM angle $\gamma =60^{\circ}$.
The dominant theoretical uncertainty comes from the hadronic parameter
$\lambda_B$ and from the variation of the renormalization scale.
%%%%%%%%%%%%%%%%%%%%%%%%%%%%%%%%%%%%%%%%%%%%%%%%%%%%%%%%%%%%%%%%%%%
% export Mathematica plot as eps file and use epstopdf to get the 
% bounding box correctly

\section{$B\to\omega\gamma$}

In this section we briefly consider the decay $\bar B^{0}\to\omega^{0}\gamma$
and discuss differences to the related mode $\bar B^{0}\to\rho^{0}\gamma$. 
We consider $\rho^{0}$ and $\omega^{0}$ as pure isospin-1 and isospin-0 modes, 
respectively, and neglect $\rho-\omega$ mixing. We use the convention 
$\omega^{0}=\frac{u\bar u+d\bar d}{\sqrt{2}}$, 
$\rho^{0}=\frac{u\bar u-d\bar d}{\sqrt{2}}$. 

To leading order in the heavy-quark limit and next-to-leading order in 
$\alpha_{s}$ both the $\rho^{0}$ and $\omega^{0}$ meson in $B\to V\gamma$ are  
produced from a $d\bar d$ pair. Therefore, to get the 
$\bar B\to\omega^{0}\gamma$ decay amplitude, we can use the one for 
$\bar B\to\rho^{0}\gamma$ with obvious replacements for the vector meson 
decay constant, mass, LCDA and form factor in the factorization coefficients 
$a_{7}^{u/c}(\rho^{0}\gamma)$ \cite{BBVgam,HL,thesis}. The relevant input 
parameters for all the decay modes are compiled in Table \ref{tab:input}.

A few comments are in order. The best known input parameter for the vector 
mesons is the mass, which can be found in the Review of Particle Physics 
\cite{PDG}. Using $\tau$-decay data and the purely leptonic decay modes of 
$\rho^{0}$ and $\omega^{0}$ one can extract the respective decay constants 
$f_{\rho,\,\omega}$ with negligible uncertainty \cite{BNPV}. The other vector 
meson parameters, such as the $B\to V$ form factors were taken from QCD 
sum rule estimates \cite{BB98}. We take the same values for the $\omega$ and 
$\rho$ mesons, which should be a
reasonable assumption, even though this equality could be broken by Zweig-rule
violating effects. The latter are, however, suppressed by $1/N_c$.
For instance, the decay constants $f_\rho$ and $f_\omega$ differ by
$10\%$.

\begin{figure}
  \begin{center}  
    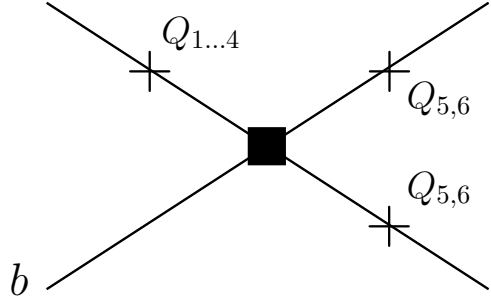
  \end{center}
\caption{Annihilation contribution to the $\bar B\to V\gamma$ decay. 
The dominant mechanism for operators $Q_{1\ldots 4}$ is the radiation of the 
photon from the light quark in the $B$ meson, as shown. This amplitude is 
suppressed by one power of $\Lambda_{QCD}/m_b$, but it is still calculable in 
QCD factorization. Radiation of the photon from the remaining three quark 
lines is suppressed by $(\Lambda_{QCD}/m_b)^2$ for operators $Q_{1\ldots 4}$. 
For operators $Q_{5,6},$ however, radiation from the final state 
quarks is again of order $\Lambda_{QCD}/m_b$.\label{fig:ann}}
\end{figure}

In \cite{BBVgam,thesis} we included weak annihilation contributions to 
$B\to V\gamma$ although they are suppressed by one power of 
$\Lambda_{\mathrm QCD}/m_{b}$. The reason for including these power-suppressed 
contributions was that they are in part enhanced by large Wilson
coefficients, they are calculable in QCD factorization and they can be 
used to estimate isospin-breaking effects. For $B\to\omega\gamma$ annihilation 
contributions are also calculable and they are the source of specific 
differences (apart from form factors) between 
$\bar B^{0}\to\rho^{0}\gamma$ and $\bar B^{0}\to\omega^{0}\gamma$. 
Those are due to the fact that $\rho^{0}$ and $\omega^{0}$ are isospin-0 and 
isospin-1 states, respectively. In the following we will use the notation of 
section 4.5 in \cite{thesis}. If, in figure~\ref{fig:ann}, 
%%%%%%%%%%%%%%%%%%%%%%%%%%%%%%%%%%%%%%%%%%%%%%%%%%%%%%%%%%%%%%%%%%%
%%%%%%%%%%%%%%%%%%%%%%%%%%%%%%%%%%%%%%%%%%%%%%%%%%%%%%%%%%%%%%%%%%%
the photon emission is from the light quark in the $B$ meson, 
the annihilation amplitude contains
\begin{equation}\label{bVdef}
  b^V =\frac{2\pi^2}{F_V}\frac{f_B m_V f_V}{m_B m_b \lambda_B}
\end{equation}
whereas the $Q_{5,6}$ insertion with the photon emitted from one of the vector 
meson constituent quarks leads to
\begin{equation}\label{dVdef}
  d^V_{\stackrel{(-)}{v}} =-\frac{4 \pi^2}{F_V}\frac{f_B f_V^\perp}{m_B m_b} 
\int_0^1 \frac{dv}{\stackrel{(-)}{v}} \Phi_V^\perp(v)
\end{equation}

The annihilation coefficients for $\bar B^{0}\to\omega^{0}\gamma$ then are
\begin{eqnarray}
\label{aann} a_{ann}^u(\omega^0\gamma)      & = & 
Q_d \left[+a_2 b^\omega -2b^{\omega}(a_{4}+a_{6}) +a_4 b^\omega +
a_6(d^\omega_v +d^\omega_{\bar v})\right]\\
a_{ann}^c(\omega^0\gamma)      & = & Q_d \left[ -2b^{\omega}(a_{4}+a_{6}) + 
a_4 b^\omega +a_6(d^\omega_v +d^\omega_{\bar v})\right]
\end{eqnarray}
The difference compared to $a_{ann}^{u/c}(\rho^0\gamma)$ is the sign change of 
the $a_{2}$ contribution and the additional isospin-0 contribution 
$-2b^{\omega}(a_{4}+a_{6})$.

Numerically the $\omega$ and $\rho^{0}$ annihilation coefficients are 
given by
\begin{eqnarray}\label{auannomeganum}
  a_{ann}^u(\omega\gamma) &=& - \frac{1}{3}\Big[
  	\begin{array}[t]{cccc} +0.0268 & +0.0281 & -0.0060 & +0.0446\\ 
	+a_2 b^{\omega} & \!\!-2b^{\omega}(a_{4}+a_{6})\!\! & +a_4 b^\omega & 
+a_6(d^\omega_v +d^\omega_{\bar v})
	\end{array}\Big]=-0.0312\\
 % &=& -0.0312\nonumber\\[0.3cm]
\label{acannrhonum}
  a_{ann}^u(\rho^{0}\gamma) &=& -\frac{1}{3}\Big[
  	\begin{array}[t]{ccc} -0.0296 & -0.0066 & +0.0446\\
	-a_2 b^{\rho} & +a_4 b^\rho & +a_6(d^\rho_v +d^\rho_{\bar v})
	\end{array}\Big] = -0.0028
\end{eqnarray}
For comparison we quote the corresponding numbers for the $\rho^-\gamma$
channel
\begin{equation}
  a_{ann}^u(\rho^{-}\gamma) = \frac{2}{3}\Big[
  	\begin{array}[t]{cccc} + 0.2902 & -0.0066 & -0.0112 & +0.0223\\
	+a_1 b^{\rho} & +a_4 b^\rho & +Q_{s}/Q_{u} a_6 d^\rho_v & 
+a_6 d^\rho_{\bar v}
	\end{array}\Big] = +0.1965
\end{equation}
where the annihilation component is considerably larger.
This has to be compared with 
$a_{7}^{u}(\omega\gamma)=a_{7}^{u}(\rho^{0}\gamma)=-0.4154 -0.0685i$,
to which the annihilation coefficients are added.
For central values of all input parameters, $\mu=m_{b}$, and our default 
choice for the CKM angle $\gamma=58^\circ$, we get the following CP-averaged 
branching ratios:
\begin{eqnarray}
\bar B(B^{0}\to\omega\gamma) &=& 0.84\cdot 10^{-6}\\
\bar B(B^{0}\to\rho^{0}\gamma) &=& 0.81\cdot 10^{-6}\\
\bar B(B^{\pm}\to\rho^{\pm}\gamma) &=& 1.81\cdot 10^{-6}
\end{eqnarray}
Within the parametric and theoretical uncertainties the $B^{0}\to\omega\gamma$ 
and $B^{0}\to\rho\gamma$ branching ratios can be considered equal,
neglecting any possile difference in the respective form factors.

\section{Conclusions and Outlook}
\label{sec:conc}

We have studied constraints on the CKM unitarity triangle from
observables in the exclusive radiative decays $B\to K^{*}\gamma$, $B
\to\rho\gamma$, and $B\to\omega\gamma$, as well as the exclusive
semileptonic decay $B\to\rho l\nu$. Within the framework of QCD
factorization we have worked at next-to-leading order in
$\alpha_{s}$ to leading order in the heavy-quark limit. 
Power corrections from weak annihilation have also been included.
Important information on the unitarity-triangle parameters 
$\bar\rho$ and $\bar\eta$ can be
obtained from the ratio $R_{0}$ of the neutral $B^{0}\to
\rho^{0}\gamma$ and $B^{0}\to K^{*0}\gamma$ branching ratios. This
ratio measures to very good approximation the side $R_{t}$ of the
standard unitarity triangle. Annihilation effects are negligible in this
case. The theoretical uncertainty in the relation to $R_t$
comes in essence solely from the form-factor ratio
$\xi=F_{K^{*}}/F_{\rho}$, which differs from unity only because of 
SU(3)-breaking effects. 
Using the latest bound on $B(B^0\to\rho^0\gamma)$ from Babar
we find $R_t < 0.81\, (\xi/1.3)$ or $|V_{td}|< 7.3\cdot 10^{-3}\, (\xi/1.3)$ 
(see also Fig. \ref{fig:R00UL}). 

Similar constraints in the
$(\bar\rho,\bar\eta)$ plane can be obtained from the isospin asymmetry
$\Delta(\rho\gamma)$ once a measurement of this quantity is available. 

We propose to gain complementary information  in the
$(\bar\rho,\bar\eta)$ plane through the $B\to\rho l\nu$ and 
$B\to\rho\gamma$ decay rates, which can be related to the CKM parameter
$|V_{ud}V_{ub}/(V_{td}V_{tb})|^{2}$. For events
where the momenta of the neutrino and the $B$ meson are parallel in the
dilepton centre-of-mass frame, this relation is free of hadronic form
factors in the large energy limit. This allows a theoretically clean
determination of the above CKM ratio.
We have shown that even a moderate experimental precision  
can yield a stringent constraint in the $(\bar\rho,\bar\eta)$ plane.

Finally, we have calculated the annihilation effects in the
$B\to\omega\gamma$ decay amplitude which turn out to be very small.

An improved determination of the $B\to V$ form
factors and, in particular, the form-factor ratio $\xi$, remains
an important task for the future.
More precise experimental measurements, specifically the
individual measurements
of $B(B^{0}\to\rho^{0}\gamma)$ and  $B(B^{+}\to\rho^{+}\gamma)$  are
eagerly awaited. These
measurements can lead to results on $R_{t}$ competitive with
those from $B_{s}$-$\bar B_{s}$ mixing.
An experimental analysis of the differential $B\to\rho l\nu$ to
$B\to\rho\gamma$ decay rate ratio can circumvent the form-factor
related uncertainties to a large extent and will thus be of
particular interest.

{\em Acknowledgements:} S.W.B. wants to thank Thorsten Feldmann for 
helpful discussion of the $B\to\omega$ form factor. This research was 
supported in part by the National Science Foundation under Grant PHY-0355005.

\end{document}